\documentclass[preprintnumbers,article,amsmath,amssymb,floatfix,10pt,prd,onecolumn,
superscriptaddress,nofootinbib]{revtex4-2}
\usepackage{bm}
\usepackage{amsfonts}
\usepackage{latexsym}
\usepackage[latin1]{inputenc}
\usepackage{graphicx}
\usepackage{amsmath}
\usepackage{palatino}
\usepackage{mathpazo}
\usepackage{textcomp}
\usepackage{lineno}
\usepackage{float}
\usepackage{booktabs}
\usepackage{dcolumn}
\usepackage{ragged2e}
\usepackage{hyperref}
\usepackage{tensor}
\usepackage[version=3]{mhchem}%to write chemical symbols
\hypersetup{colorlinks,citecolor=blue}
\hypersetup{colorlinks=true,linkcolor=red,filecolor=magenta,    urlcolor=blue}
\usepackage{amsmath}
\usepackage{xcolor}
\usepackage{enumitem}
\usepackage{orcidlink}
\usepackage{epsfig}
\usepackage{subfigure}
\usepackage{commath}
\usepackage{caption}
\def\jnl@style{\it}
\def\aaref@jnl#1{{\jnl@style#1}}

\def\aaref@jnl#1{{\jnl@style#1}}

\def\aj{\aaref@jnl{AJ}}                   % Astronomical Journal
\def\apj{\aaref@jnl{ApJ}}                 % Astrophysical Journal
\def\apjl{\aaref@jnl{ApJ}}                % Astrophysical Journal, Letters
\def\apjs{\aaref@jnl{ApJS}}               % Astrophysical Journal, Supplement
\def\apss{\aaref@jnl{Ap\&SS}}             % Astrophysics and Space Science
\def\aap{\aaref@jnl{A\&A}}                % Astronomy and Astrophysics
\def\aapr{\aaref@jnl{A\&A~Rev.}}          % Astronomy and Astrophysics Reviews
\def\aaps{\aaref@jnl{A\&AS}}              % Astronomy and Astrophysics, Supplement
\def\mnras{\aaref@jnl{Mon.~Not.~Roy.~Astron.~Soc.}}             % Monthly Notices of the RAS
\def\prd{\aaref@jnl{Phys.~Rev.~D}}        % Physical Review D
\def\prc{\aaref@jnl{Phys.~Rev.~C}}  % Physical Review C
\def\prl{\aaref@jnl{Phys.~Rev.~Lett.}}    % Physical Review Letters
\def\qjras{\aaref@jnl{QJRAS}}             % Quarterly Journal of the RAS
\def\skytel{\aaref@jnl{S\&T}}             % Sky and Telescope
\def\ssr{\aaref@jnl{Space~Sci.~Rev.}}     % Space Science Reviews
\def\zap{\aaref@jnl{ZAp}}                 % Zeitschrift fuer Astrophysik
\def\nat{\aaref@jnl{Nature}}              % Nature
\def\aplett{\aaref@jnl{Astrophys.~Lett.}} % Astrophysics Letters
\def\apspr{\aaref@jnl{Astrophys.~Space~Phys.~Res.}} % Astrophysics Space Physics Research
\def\physrep{\aaref@jnl{Phys.~Rep.}}      % Physics Reports
\def\physscr{\aaref@jnl{Phys.~Scr}}       % Physica Scripta
\def\commat{\aaref@jnl{Comm.~Math.~Phys.}}              % Communications in Mathematical Physics
\def\science{\aaref@jnl{Science}}               % Science
\def\cqg{\aaref@jnl{Classical Quant.~Grav.}}            % Classical and Quantum Gravity
\def\jpcs{\aaref@jnl{JPCS}}                                     % Journal of Physics Conference Series
\def\ijmpd{\aaref@jnl{Int.~J.~Mod.~Phys.~D}}                    % International Journal of Modern Physics D
\def\grg{\aaref@jnl{Gen.~Relat.~Gravit.}}               % General Relativity and Gravitation
\def\rpp{\aaref@jnl{Rep.~Prog.~Phys.}}          % Reports on Progress in Physics
\def\npa{\aaref@jnl{Nucl.~Phys.~A}}        % Nuclear Physics A
\def\lrr{\aaref@jnl{Living Rev.~Rel.}}                   % Living reviews in relativity
\def\jcap{\aaref@jnl{J.~Cosmology Astropart.~Phys.}}    % Journal of cosmology and astroparticle physics
\def\rmp{\aaref@jnl{Rev.~Mod.~Phys.}}   %Reviews of modern physics
\def\epjc{\aaref@jnl{Eur.~Phys.~J.~C}} 
\def\plb{\aaref@jnl{~Phy.~Lett.~B}} 
\def\mpla{\aaref@jnl{Mod.~Phy.~Lett.~A}} 
\def\arxiv{\aaref@jnl{arxiv.org}}

\newcommand{\unit}[1]{\ensuremath{\, \mathrm{#1}}}
\allowdisplaybreaks[1]
\begin{document}

\title{Perturbation Dynamics and Structure Formation in Extended Proca-Nuevo Gravity}

\author{N. S. Kavya\orcidlink{0000-0001-8561-130X}}
\email{kavya.samak.10@gmail.com}
\affiliation{Institute of Mathematical Sciences, Faculty of Science, Universiti Malaya, 50603 Kuala Lumpur, Malaysia}%

\author{Avik De\orcidlink{0000-0001-6475-3085}}
\email{avikde@um.edu.my}
\affiliation{Institute of Mathematical Sciences, Faculty of Science, Universiti Malaya, 50603 Kuala Lumpur, Malaysia}%

\author{Tee-How Loo\orcidlink{0000-0003-4099-9843}}
\email{looth@um.edu.my}
\affiliation{Institute of Mathematical Sciences, Faculty of Science, Universiti Malaya, 50603 Kuala Lumpur, Malaysia}%

\begin{abstract}
A comprehensive analysis of cosmological perturbations and structure formation is presented for the Extended Proca-Nuevo (EPN) framework, a vector-tensor extension of General Relativity with a massive spin-1 field. In this scenario, the vector field modifies the background expansion through an algebraic constraint, leading to a characteristic Hubble evolution that interpolates between $\Lambda$CDM limits while introducing deviations at intermediate redshifts. Assuming minimal matter coupling, the linear perturbation equations are derived in gauge-invariant form and the resulting growth of matter inhomogeneities is analyzed. The EPN sector induces effective anisotropic stress and couples a single propagating scalar mode to the metric potentials, leaving the matter growth equation in its GR form but with a modified expansion history. The full scalar perturbation system is presented, stability conditions are discussed, and the results provide a foundation for testing the EPN framework against current and future cosmological observations.
\end{abstract}

\footnotetext{The research was supported by the Ministry of Higher Education (MoHE), through the Fundamental Research Grant Scheme (FRGS/1/2023/STG07/UM/02/3, project no.: FP074-2023).}

\maketitle

\section{Introduction}\label{sec:introEPNTgrowth}

The era of precision cosmology has provided unprecedented insight into the dynamical history of the Universe, allowing for the systematic reconstruction of its expansion as well as the formation and evolution of cosmic structures across a wide range of scales. Within the standard $\Lambda$CDM framework, the late-time acceleration of the Universe is attributed to a cosmological constant, while General Relativity (GR) governs the evolution of both the homogeneous background and linear cosmological perturbations. This paradigm has demonstrated remarkable success in describing observations from the cosmic microwave background (CMB), large-scale structure (LSS), baryon acoustic oscillations, and weak gravitational lensing. However, despite its empirical effectiveness, $\Lambda$CDM remains a phenomenological framework whose underlying physical interpretation is incomplete. In particular, the cosmological constant lacks a compelling theoretical origin and suffers from well-known conceptual issues, such as the extreme mismatch between the observed vacuum energy density and the natural value predicted by quantum field theory, as well as the coincidence problem, namely why the dark energy density becomes dynamically relevant only in the recent cosmological epoch~\citep{Weinberg:1988cp,Padmanabhan:2002ji, Zlatev:1998tr}.

Beyond these theoretical challenges, the standard model of cosmology treats dark energy as an exactly homogeneous and non-clustering component, implying that its role is restricted to modifying the background expansion while leaving gravitational clustering entirely governed by cold dark matter and GR. Although this assumption leads to a remarkably successful phenomenological description, it significantly limits the range of dynamical possibilities in the late Universe and obscures the potential role of additional gravitational degrees of freedom. Furthermore, as cosmological measurements continue to increase in precision, subtle deviations in structure growth, gravitational interaction strengths, and perturbation dynamics provide growing motivation to explore frameworks in which the dark sector and gravity may interact in richer ways than permitted within $\Lambda$CDM~\citep{Choudhury:2025bnx,Lesgourgues:2015wza,Joyce:2014kja}. In this sense, the search for theoretically consistent extensions of GR capable of altering both the expansion history and the perturbative evolution of cosmic structures represents a natural and timely direction in modern cosmology.

From a theoretical standpoint, the study of structure formation and perturbation dynamics offers one of the most powerful avenues for testing gravitational theories beyond GR. While distinct models may yield similar expansion histories at the background level, their predictions for the growth of matter perturbations, gravitational potentials, and effective dark-sector clustering can differ significantly. The growth rate of cosmic structures, the evolution of metric perturbations, and the emergence of effective anisotropic stresses therefore serve as key discriminators among competing frameworks~\citep{Giovannini:2011tk,Pogosian:2007sw,Jain:2007yk}. A viable modified-gravity theory must not only reproduce a realistic late-time expansion but must also yield a perturbation sector that is dynamically stable, free of pathological modes, and consistent with present observational constraints across both linear and quasi-linear scales. The search for theoretically consistent extensions of GR has led to diverse frameworks, among which vector--tensor theories have emerged as particularly compelling scenarios~\citep{Milgrom:2009gv,deRham:2010kj,Comelli:2011zm,Eling:2004dk,DeFelice:2016yws}. These theories offer rich dynamical structure while maintaining consistency through carefully designed constraint mechanisms, with massive spin-1 fields providing novel avenues for modifying gravitational interactions. Within this context, the Proca-Nuevo (PN) and Extended Proca-Nuevo (EPN) frameworks provide a novel realization of a massive vector field with nontrivial self-interactions and derivative couplings that preserve the correct number of propagating degrees of freedom and avoid ghost-like instabilities \citep{deRham:2020yet,ErrastiDiez:2023gme, Saridakis:2021qxb, Sudharani:2024qnn,Sultan:2025dqr,Kavya:2025gfw,Chiang:2025hrj}. These constructions enable the vector field to play an active cosmological role while maintaining mathematical consistency and predictivity.

A distinctive feature of the EPN framework is that its cosmological background dynamics give rise to an emergent effective dark-energy sector whose behavior is governed by an algebraic constraint linking the vector field profile to the Hubble expansion. Unlike scalar-tensor extensions, where additional fields typically introduce new propagating dynamical degrees of freedom at the background level, the EPN theory modifies the cosmological evolution through non-canonical powers of the Hubble parameter, while preserving a constrained dynamical structure. This mechanism produces characteristic departures from $\Lambda$CDM at intermediate epochs while recovering stable behavior in the asymptotic limits, thereby offering a theoretically controlled setting in which deviations from standard gravity may naturally arise~\citep{Anagnostopoulos:2023pvi}. These departures motivate a detailed investigation of how the EPN framework influences the formation and growth of cosmic structures.

In the present work, our interest lies in examining cosmic structure growth and perturbation dynamics within the EPN framework. While earlier analyses have primarily focused on background evolution and the emergence of effective late-time acceleration, a comprehensive assessment of linear perturbations is essential for establishing the phenomenological viability of the theory. The analysis of scalar, vector, and tensor perturbations provides insight into how the massive vector field affects gravitational clustering, how effective gravitational couplings evolve in time, and whether departures from the standard growth history arise in a dynamically consistent manner. In particular, one may follow the evolution of the matter density contrast, the effective growth rate, and the metric potentials associated with the EPN interaction structure, thereby revealing possible signatures that distinguish the theory from GR-based cosmology. A central motivation for this exploration is that perturbation observables place stringent constraints on any modified-gravity scenario. The presence or absence of anisotropic stress, the behavior of subhorizon scalar modes, and the stability properties of the perturbative degrees of freedom play a decisive role in determining whether an effective dark-energy sector clusters or remains smooth. Vector-tensor theories, by virtue of their intrinsic spin-1 structure, enrich the perturbation space relative to scalar-tensor models, introducing new couplings and potential phenomenological features \citep{Lagos:2016wyv,Lagos:2017hdr,Aoki:2021wew,Heisenberg:2018acv,Kase:2018nwt}. The EPN framework, with its constrained vector dynamics and algebraic background relations, offers a particularly well-defined arena in which to examine these effects. This work develops a systematic investigation of cosmic structure growth and perturbation evolution in the EPN framework. Our objective is to clarify how the underlying vector self-interactions and constraint structure manifest in the cosmological perturbation sector, and to characterize the resulting signatures at both the background and linear levels.

The manuscript is organized as follows: In Section~\ref{sec:epntcosmo}, we review the theoretical formulation of the covariant EPN theory in a cosmological setting and summarize the background evolution equations that arise from the interplay between the vector field and gravity. In Section~\ref{sec:growthEPN}, we analyze the perturbation structure and the resulting growth of matter in this framework, emphasizing the implications of the vector degrees of freedom for the dynamics of structure formation. Section~\ref{sec:observational} describes our observational methodology, including the datasets and Bayesian analysis techniques used to confront the EPN framework with current cosmological measurements. In Section~\ref{sec:results}, we present the parameter constraints and assess the quality of fits to various cosmological probes. Section~\ref{sec:conclusion} discusses the theoretical implications of our findings and outlines directions for future research. 

\section{Cosmology of the Extended Proca-Nuevo Theory}\label{sec:epntcosmo}

We consider the covariant realization of the EPN framework minimally coupled to Einstein gravity, in which a massive spin-1 field endowed with nonlinear self-interactions contributes dynamically to the cosmic expansion. The action is given by
\begin{equation}\label{eq:action_epn}
\mathcal{S}
=\int d^4x\,\sqrt{-g}
\left[
\frac{M_{\mathrm{Pl}}^2}{2}R
+\mathcal{L}_{\mathrm{EPN}}
+\mathcal{L}_{m}
\right],
\end{equation}
where $R$ is the Ricci scalar, $\mathcal{L}_{m}$ denotes the matter Lagrangian, and $\mathcal{L}_{\mathrm{EPN}}$ encodes the dynamics of the massive vector field and its derivative interactions~\citep{deRham:2021efp,deRham:2023brw}.

The EPN Lagrangian takes the form
\begin{equation}\label{eq:EPNLagrangian}
\mathcal{L}_{\mathrm{EPN}}
=-\frac{1}{4}\mathcal{F}^{\mu\nu}\mathcal{F}_{\mu\nu}
+\Lambda^{4}\sum_{n=0}^{3}\mathcal{L}_{n},
\end{equation}
where $\mathcal{F}_{\mu\nu}
=\nabla_{\mu}\mathcal{V}_{\nu}-\nabla_{\nu}\mathcal{V}_{\mu}$ is the field-strength tensor, $\Lambda$ is the interaction scale, and the operators $\mathcal{L}_{n}$ are constructed from the tensor. The interaction operators $\mathcal{L}_n$ are constructed as the elementary symmetric polynomials of the tensor $\mathcal{K}^{\mu}{}_{\nu}$, given by
\begin{align}
\mathcal{L}_0 &= 1, \\
\mathcal{L}_1 &= [\mathcal{K}], \\
\mathcal{L}_2 &= [\mathcal{K}]^2 - [\mathcal{K}^2], \\
\mathcal{L}_3 &= [\mathcal{K}]^3 - 3[\mathcal{K}][\mathcal{K}^2] + 2[\mathcal{K}^3],
\end{align}
where square brackets denote traces, e.g. $[\mathcal{K}] = \mathcal{K}^\mu_{\ \mu}$. These combinations ensure that the theory propagates the correct number of degrees of freedom and remains free from ghost instabilities.

\begin{equation}\label{eq:Ktensor_epn}
\mathcal{K}^{\mu}{}_{\nu}
=\mathcal{X}^{\mu}{}_{\nu}
-\delta^{\mu}{}_{\nu},\qquad
\mathcal{X}^{\mu}{}_{\nu}
=\left(\sqrt{\eta^{-1}f[\mathcal{V}]}\right)^{\mu}{}_{\nu},
\end{equation}

\noindent
with the 
St\"uckelberg-type tensor
\begin{equation}\label{eq:fmndef}
f_{\mu\nu}
=\eta_{\mu\nu}
+2\frac{\partial_{(\mu}\mathcal{V}_{\nu)}}{\Lambda^{2}}
+\frac{\partial_{\mu}\mathcal{V}^{\rho}\partial_{\nu}\mathcal{V}_{\rho}}{\Lambda^{4}}.
\end{equation}
We note that the derivatives appearing in Eq.~\eqref{eq:fmndef} are partial derivatives, consistent with the St\"uckelberg construction of the Proca-Nuevo framework. The tensor $f_{\mu\nu}$ is an auxiliary object and does not correspond to a covariant spacetime tensor, hence the use of $\partial_\mu$ rather than $\nabla_\mu$.

\noindent
The interaction functions depend on the invariant
\begin{equation}\label{eq:Xinv}
X=-\frac{1}{2\Lambda^{2}}
\,\mathcal{V}^{\mu}\mathcal{V}_{\mu},
\end{equation}
ensuring the correct number of propagating degrees of freedom and the absence of ghostlike instabilities~\citep{deRham:2021efp}.

We now specialize to a spatially flat Friedmann-Lema\^itre-Robertson-Walker background,
\begin{equation}\label{eq:FLRWmetric}
ds^{2}
=-dt^{2}
+a^{2}(t)\,\delta_{ij}\,dx^{i}dx^{j},
\end{equation}
and adopt a homogeneous, isotropic vector-field configuration,
\begin{equation}\label{eq:vecprofile_epn}
\mathcal{V}_{\mu}dx^{\mu}
=-\phi(t)\,dt.
\end{equation}

Here, in Eq. \eqref{eq:Ktensor_epn}, $\eta_{\mu\nu}$ denotes the reference metric entering the St\"uckelberg construction of the Proca-Nuevo theory. This does not correspond to the physical spacetime metric, which is given by $g_{\mu\nu}$ in Eq.~\eqref{eq:FLRWmetric}. The square-root structure therefore remains well-defined in a fully covariant setting.

Variation of~\eqref{eq:action_epn} with respect to the metric leads to the modified Friedmann equations
\begin{gather}
H^{2}
=\frac{1}{3M_{\mathrm{Pl}}^{2}}
\left(\rho_{m}+ {\rho}_{\mathrm{EPN}}\right),\label{eq:Fried1}\\
\dot{H}+H^{2}
=-\frac{1}{6M_{\mathrm{Pl}}^{2}}
\left(\rho_{m}+ {\rho}_{\mathrm{EPN}}
+3p_{m}+3 {p}_{\mathrm{EPN}}\right),\label{eq:Fried2}
\end{gather}
where $( {\rho}_{\mathrm{EPN}}, {p}_{\mathrm{EPN}})$ represent the effective dark-energy sector emerging from the massive vector field.

The variation with respect to $\phi(t)$ yields a non-dynamical constraint,
\begin{equation}\label{eq:constraint_epn}
\alpha_{0,X}
+3\left(\alpha_{1,X}+d_{1,X}\right)
\frac{H\phi}{\Lambda^{2}}=0,
\end{equation}
which enforces an algebraic relation between $H$ and $\phi$, thereby eliminating unwanted extra degrees of freedom.

The effective dark-energy density and pressure take the form
\begin{gather}
 {\rho}_{\mathrm{EPN}}
=\Lambda^{4}
\left[
-\alpha_{0}
+\alpha_{0,X}\frac{\phi^{2}}{\Lambda^{2}}
+3\left(\alpha_{1,X}+d_{1,X}\right)
\frac{H\phi^{3}}{\Lambda^{4}}
\right],\label{eq:rhoE_epn}\\
 {p}_{\mathrm{EPN}}
=\Lambda^{4}
\left[
\alpha_{0}
-\left(\alpha_{1,X}+d_{1,X}\right)
\frac{\phi^{2}\dot{\phi}}{\Lambda^{4}}
\right].\label{eq:pE_epn}
\end{gather}
Using~\eqref{eq:constraint_epn}, the energy density can be recast as
\begin{equation}\label{eq:rhoEcompact_epn}
 {\rho}_{\mathrm{EPN}}
=\Lambda^{4}\frac{c_{m}y^{2/3}}{2}
\left(
\frac{\Lambda^{4}}{M_{\mathrm{Pl}}^{2}H^{2}}
\right)^{1/3},
\end{equation}
where $c_{m}\sim \mathcal{O}(1)$ and $y$ are constants determined by the interaction structure~\citep{deRham:2020yet}. Substituting~\eqref{eq:rhoEcompact_epn} into~\eqref{eq:Fried1} yields the algebraic Hubble evolution equation
\begin{equation}
\label{eq:HubbleAlgebraic}
H^{2}
-H_{0}^{2}\Omega_{m0}(1+z)^{3}
-(1-\Omega_{m0})H_{0}^{8/3}H^{-2/3}=0,
\end{equation}
which exhibits the characteristic powered-Hubble correction of EPN cosmology.

The presence of the $H^{-2/3}$ term distinguishes EPN dynamics from $\Lambda$CDM while reproducing the standard limits at very early and very late epochs. Deviations emerge predominantly at intermediate redshifts, precisely where observational cosmological dynamics are most pronounced, making the EPN framework a natural and theoretically consistent arena in which to reassess the interpretation of late-time observables and their connection to early-Universe inferences.
\bigskip

\section{Matter Growth in EPN Cosmology: Minimal Coupling and GR-Form Growth Equation}\label{sec:growthEPN}

In the present framework, the matter sector is assumed to be minimally coupled to Einstein gravity, while the EPN vector field contributes only to the homogeneous background evolution without introducing new clustering degrees of freedom at the level of sub-horizon matter perturbations. In particular, the EPN field does not modify the Poisson equation, does not mediate an effective fifth force, and does not introduce scale-dependent corrections to the growth of cold dark matter perturbations. Consequently, the evolution of the linear matter density contrast retains the same dynamical structure as in GR, with the only modification entering through the non-standard background expansion history $H(a)$ associated with the EPN cosmology~\citep{Ruiz-Zapatero:2022xbv,Archidiacono:2022iuu,Saridakis:2021qxb}.

Under these assumptions, the linear growth of pressureless matter perturbations in the sub-horizon and quasi-static regime is governed by the standard GR growth equation written in terms of the scale factor $a$,
\begin{equation}
\delta''(a)
+\left(\frac{3}{a}
+\frac{H'(a)}{H(a)}\right)\delta'(a)
-\frac{3\,\Omega_{m0}H_{0}^{2}}{2a^{5}H^{2}(a)}\,
\delta(a)=0,
\label{eq:growthGRform}
\end{equation}
where a prime denotes differentiation with respect to $a$, $H(a)$ is the background Hubble function, and $\Omega_{m0}$ is the present matter density parameter. The structure of Eq.~\eqref{eq:growthGRform} reflects the fact that the gravitational coupling remains unmodified, i.e.\ $G_{\rm eff}=G$, and therefore the source term driving structure formation is identical to that of the $\Lambda$CDM scenario, with the difference that the expansion rate entering the friction and source terms corresponds to the EPN background rather than to the standard one~\citep{Alho:2019jho,Rampf:2022tpg,Kavya:2025vsj,Mishra:2025kzu,Sudharani:2025cii,Mishra:2025rhi,Li:2025msm,Tzerefos:2023hpi,Subramaniam:2024uuu}.

The key task is therefore to compute the derivative $H'(a)$ consistently from the implicit algebraic background equation that characterizes the EPN cosmology. The background expansion history is determined by the algebraic relation
\begin{equation}
H^{2}
-H_{0}^{2}\Omega_{m0}a^{-3}
-(1-\Omega_{m0})H_{0}^{8/3}\,H^{-2/3}=0,
\label{eq:EPNequation}
\end{equation}
which expresses the effective dark-energy contribution as a non-canonical, powered function of the Hubble parameter. Since Eq.~\eqref{eq:EPNequation} is implicit in $H(a)$, the derivative $H'(a)$ must be obtained by implicit differentiation.

To proceed, we define
\begin{equation}
A_0 \equiv H_0^{2}\Omega_{m0},
\qquad
B_0 \equiv (1-\Omega_{m0})H_{0}^{8/3},
\end{equation}
so that Eq.~\eqref{eq:EPNequation} may be written compactly as
\begin{equation}
F(H,a)=H^{2}-A_0a^{-3}-B_0H^{-2/3}=0.
\end{equation}
Differentiating with respect to the scale factor and invoking the implicit function theorem gives
\begin{equation}
\frac{\partial F}{\partial a}
+\frac{\partial F}{\partial H}\,H'(a)=0,
\end{equation}
from which the relevant derivatives are
\begin{equation}
\frac{\partial F}{\partial a}
=+3A_0a^{-4},
\qquad
\frac{\partial F}{\partial H}
=2H+\frac{2}{3}B_0H^{-5/3}.
\end{equation}
Solving for $H'(a)$ yields
\begin{equation}
H'(a) 
=-\frac{3A_0a^{-4}}
{2H+\frac{2}{3}B_0H^{-5/3}},
\label{eq:Hprime}
\end{equation}
which represents the consistent first derivative of the EPN Hubble function derived directly from the algebraic constraint. For convenience, one may also express the logarithmic derivative
\begin{equation}
\frac{H'(a)}{H(a)}
=-\frac{3A_0a^{-4}}
{H\!\left(2H+\frac{2}{3}B_0H^{-5/3}\right)},
\label{eq:Hratio}
\end{equation}
which enters explicitly in the friction term of the growth equation.

Equations~\eqref{eq:growthGRform} and~\eqref{eq:Hratio}, together with the background relation~\eqref{eq:EPNequation}, fully determine the growth of matter perturbations in the EPN scenario in the sub-horizon regime, under the assumption of minimal matter coupling and a non-clustering EPN sector. The resulting phenomenology differs from $\Lambda$CDM exclusively through the modified expansion history, rather than through a change in the effective gravitational coupling or the introduction of new propagating perturbation modes in the matter sector.

Having established the sub-horizon growth behaviour under the minimal-coupling assumption, we now proceed to the full gauge-invariant perturbation framework, where the EPN sector contributes dynamically to the scalar perturbation equations. We begin with scalar perturbations of the FLRW metric in an arbitrary gauge,
\begin{equation}
ds^2 = -(1+2A)\,dt^2
+2a\,\partial_i B\,dt\,dx^i
+a^2\!\left[(1-2\psi)\delta_{ij}
+2\partial_i\partial_jE\right]dx^i dx^j ,
\end{equation}
and construct the gauge-invariant Bardeen potentials,
\begin{equation}
\Phi \equiv A-\frac{d}{dt}\!\left[a^2(\dot E-B/a)\right],
\qquad
\Psi \equiv \psi + H\left[a^2(\dot E-B/a)\right],
\end{equation}
which encode the physical scalar-gravity degrees of freedom in a gauge-independent way~\citep{Gumrukcuoglu:2011zh,Thaalba:2024crk,Gao:2020qxy}.

The background vector configuration takes the purely temporal form
\begin{equation}
\mathcal{V}_\mu = (-\phi(t),0,0,0),
\end{equation}
and we introduce scalar perturbations via
\begin{equation}
\delta\mathcal{V}_0=-\delta\phi,
\qquad
\delta\mathcal{V}_i=\partial_i\chi .
\end{equation}
To ensure St\"uckelberg safety, we define the gauge-invariant vector-field combinations
\begin{equation}
\hat{\phi} \equiv \delta\phi-\dot\phi\,\xi^0,
\qquad
\hat{\chi}\equiv \chi-\phi\,\xi ,
\end{equation}
where $(\xi^0,\xi)$ generate scalar gauge transformations. All perturbation equations are therefore written in terms of 
$(\Phi,\Psi,\hat{\phi},\hat{\chi})$.

At the level of the perturbed action, the EPN sector behaves as an effective fluid with background quantities $( \rho_{\rm EPN},  p_{\rm EPN})$ and perturbations
\begin{equation}
\delta \rho_{\rm EPN},\qquad
\delta  p_{\rm EPN},\qquad
  q_{\rm EPN},\qquad
 \pi_{\rm EPN},
\end{equation}
obtained from variations of $\mathcal{L}_{\rm EPN}$ with respect to $g_{\mu\nu}$ and the perturbed vector field. Their explicit form depends on $\alpha_0(X)$, $\alpha_1(X)$, $d_1(X)$ and their $X$--derivatives.

A crucial structural feature is the persistence of the background algebraic constraint
\begin{equation}
\alpha_{0,X}
+3(\alpha_{1,X}+d_{1,X})\frac{H\phi}{\Lambda^2}=0 ,
\label{eq:constraint_epn_pert}
\end{equation}
which removes the would-be longitudinal ghost mode. At linear order, this constraint appears as an algebraic relation among $(\hat{\phi},\hat{\chi},\Phi,\Psi)$, implying that one scalar combination of the vector perturbations is non-dynamical. As a consequence, only a single propagating scalar degree of freedom remains, ensuring that the perturbation sector is second-order and ghost-free~\citep{Lagos:2016wyv,Kobayashi:2012kh}.

The gauge-invariant Einstein equations in Fourier space are given by
\begin{align}
-3H(\dot\Psi+H\Phi)+\frac{k^2}{a^2}\Psi
&=4\pi G\left(\delta\rho_m+\delta \rho_{\rm EPN}\right),
\label{eq:E1}\\[4pt]
\dot\Psi+H\Phi
&=4\pi G\left(\rho_m v_m+  q_{\rm EPN}\right),
\label{eq:E2}\\[4pt]
\ddot\Psi+H(\dot\Phi+2\dot\Psi)
+(2\dot H+3H^2)\Phi
&=4\pi G\left(\delta p_m+\delta  p_{\rm EPN}\right),
\label{eq:E3}\\[4pt]
\Phi-\Psi
&=8\pi G\, \pi_{\rm EPN},
\label{eq:E4}
\end{align}
so that, unlike $\Lambda$CDM, EPN generically induces anisotropic stress and hence $\Phi\neq\Psi$.

Because matter remains minimally coupled, its conservation equations retain their GR form,
\begin{equation}
\dot{\delta}_m=-\frac{k^2}{a^2}v_m+3\dot\Psi,
\qquad
\dot v_m+Hv_m=\Phi ,
\label{eq:M1-M2}
\end{equation}
while all modifications arise through the EPN sector.

Variation of the action with respect to $\mathcal{V}_\mu$ yields two scalar perturbation equations: an algebraic longitudinal constraint
\begin{equation}
\mathcal{C}_0[\hat{\phi},\hat{\chi},\Phi,\Psi]=0 ,
\label{eq:V1}
\end{equation}
which is used to eliminate one vector-field perturbation variable, and a dynamical spatial-longitudinal equation
\begin{equation}
\mathcal{D}\,\ddot{\hat{\chi}}
+\mathcal{E}\,\dot{\hat{\chi}}
+\mathcal{F}\,\hat{\chi}
=
\mathcal{S}_\Phi\,\Phi
+\mathcal{S}_\Psi\,\Psi
+\mathcal{S}_k\,\frac{k^2}{a^2}\hat{\chi},
\label{eq:V2}
\end{equation}
where the coefficients depend on $(H,\phi,X)$ and on the interaction functions and their derivatives.

Taken together, Eqs.~\eqref{eq:E1}--\eqref{eq:E4}, \eqref{eq:M1-M2}, and \eqref{eq:V1}--\eqref{eq:V2} form a closed, gauge-invariant perturbation system describing the scalar sector of EPN cosmology. This formulation makes explicit where departures from GR arise through anisotropic stress, effective-fluid source terms, and the coupling of the single dynamical EPN scalar mode to the metric potentials and provides the foundation for our analysis of structure growth and perturbation stability in the EPN framework~\citep{Perenon:2015sla}.

This assumption applies strictly within the quasi-static, sub-horizon regime considered above; in the full gauge-invariant perturbation framework, the EPN sector may contribute through anisotropic stress and through the dynamics of the surviving scalar vector mode. If the EPN interactions were to lead to an effective scale-dependent gravitational coupling, to anisotropic stress, or to an additional propagating scalar degree of freedom, then the source term in Eq.~\eqref{eq:growthGRform} would be replaced by $-(3/2)\,\Omega_{m}(a)\,G_{\rm eff}(a,k)\,\delta(a)$, and the full perturbation system would need to be re-derived from the coupled metric-vector perturbation equations. In the present analysis, however, we restrict attention to the minimally coupled case, in which the growth equation retains its GR form and the EPN sector manifests solely through its impact on the background expansion.

\subsection{Growth Factor and Observable $f\sigma_8$}
\label{subsec:growthfactor}

To establish a direct connection between the theoretical growth of matter perturbations and observations, it is convenient to introduce the linear growth factor $D(a)$, defined as
\begin{equation}
D(a) \equiv \frac{\delta(a)}{\delta(a_i)},
\end{equation}
where $a_i$ is an initial scale factor chosen deep in the matter-dominated era. The normalization is fixed such that $D(a_i)=1$, ensuring that $D(a)$ captures only the relative growth of structures.

A key quantity probed by redshift-space distortion (RSD) measurements is the logarithmic growth rate
\begin{equation}
f(a) \equiv \frac{d\ln D}{d\ln a}
= \frac{a}{\delta(a)}\,\frac{d\delta(a)}{da}.
\end{equation}
This quantity measures the rate at which matter perturbations grow with cosmic expansion and provides a sensitive test of cosmological dynamics.

The primary observable used in large-scale structure surveys is the combination
\begin{equation}
f\sigma_8(a) = f(a)\,\sigma_8(a),
\end{equation}
where $\sigma_8(a)$ denotes the root-mean-square amplitude of matter fluctuations on scales of $8\,h^{-1}\mathrm{Mpc}$. Its evolution is related to the growth factor via
\begin{equation}
\sigma_8(a) = \sigma_{8,0}\,D(a),
\end{equation}
with $\sigma_{8,0}$ being the present-day value.

In the EPN framework considered here, the evolution of $\delta(a)$ is governed by the GR-form growth equation~\eqref{eq:growthGRform}, and therefore the definitions above follow directly without modification. The impact of the EPN sector enters exclusively through the background expansion rate $H(a)$, which alters the growth history indirectly. As a result, deviations from $\Lambda$CDM in the observable $f\sigma_8(a)$ arise purely due to the modified expansion dynamics, providing a clean and robust probe of the theory.
At early times, the growth factor approaches the standard matter-dominated behavior $D(a)\propto a$, which we use to set initial conditions for numerical integration.

\section{Observational Methodology and Data}\label{sec:observational}

Modern cosmology has entered a precision-driven era in which the quantitative confrontation between theory and an unprecedented diversity of high-accuracy observational datasets has become central to establishing predictive and falsifiable models of the Universe \citep{Saridakis:2025ltr}. To confront the EPN framework with observations, we perform a Bayesian parameter inference analysis using complementary datasets that probe both the background expansion and the growth of cosmic structure. The characteristic $H^{-2/3}$ correction to the Hubble evolution~\eqref{eq:HubbleAlgebraic} produces deviations from $\Lambda$CDM predominantly at intermediate redshifts, making it essential to combine geometrical and growth-sensitive probes. Parameter estimation is carried out using a Markov Chain Monte Carlo (MCMC) analysis implemented with the {\tt emcee} affine-invariant sampler~\citep{Foreman-Mackey:2012any}.

We consider the parameter set $\{H_0,\; r_d,\; S_8,\; \Omega_{m0}\}$,
where $H_0$ and $\Omega_{m0}$ determine the background expansion, $r_d$ calibrates the BAO scale, and $S_8=\sigma_8(\Omega_m/0.3)^{0.5}$ encodes the amplitude of late-time clustering. In the minimally coupled EPN scenario the vector sector is fixed by the algebraic constraint~\eqref{eq:constraint_epn}, so no additional cosmological parameters are introduced beyond those of flat $\Lambda$CDM. We adopt broad, physically motivated priors spanning current observational ranges.

Here, we employ five independent datasets providing complementary constraints. DESI DR2 BAO measurements across \(0.295 \leq z \leq 2.33\) constrain $D_V(z)/r_d$ and its anisotropic components~\citep{Hahn:2022dnf,Raichoor:2022jab,DESI:2022gle,Chaussidon:2022pqg,DESI:2025zgx}. Type Ia supernova distances are incorporated through the Pantheon+SH0ES sample~\citep{Perivolaropoulos:2023iqj,Brout:2022vxf}, with Union3.0 used as an independent cross-check. Direct expansion-rate measurements are included via cosmic chronometers $H(z)$ determinations over $0.07<z<1.97$~\citep{Moresco:2012jh, Moresco:2015cya, Moresco:2016mzx, Jimenez:2003iv,Moresco:2020fbm}. Finally, redshift-space distortion data \citep{Alestas:2022gcg} constrain the growth rate through $f\sigma_8(z)$, computed by solving the growth equation~\eqref{eq:growthGRform}.

For datasets DESI DR2,  CC (non-correlated), and RSD we adopt a standard $\chi^2$ likelihood, while supernova likelihoods and correlated CC are evaluated using the full covariance matrices. The total likelihood is the sum of the individual contributions. For each parameter sample, we (i) solve the algebraic background relation~\eqref{eq:HubbleAlgebraic} to obtain $H(z)$, (ii) compute distance measures by numerical integration of $1/H(z)$, and (iii) integrate Eq.~\eqref{eq:growthGRform} to obtain $\delta_m(z)$ and $f\sigma_8(z)$, with $\sigma_8(z)$ evolved from $S_8$. This multi-probe strategy enables a consistent assessment of the EPN framework across both expansion and growth observables, allowing us to test whether its distinctive powered-Hubble dynamics can reconcile late-time structure measurements with background constraints.

\begin{figure}
    \centering
    \includegraphics[width=0.7\linewidth]{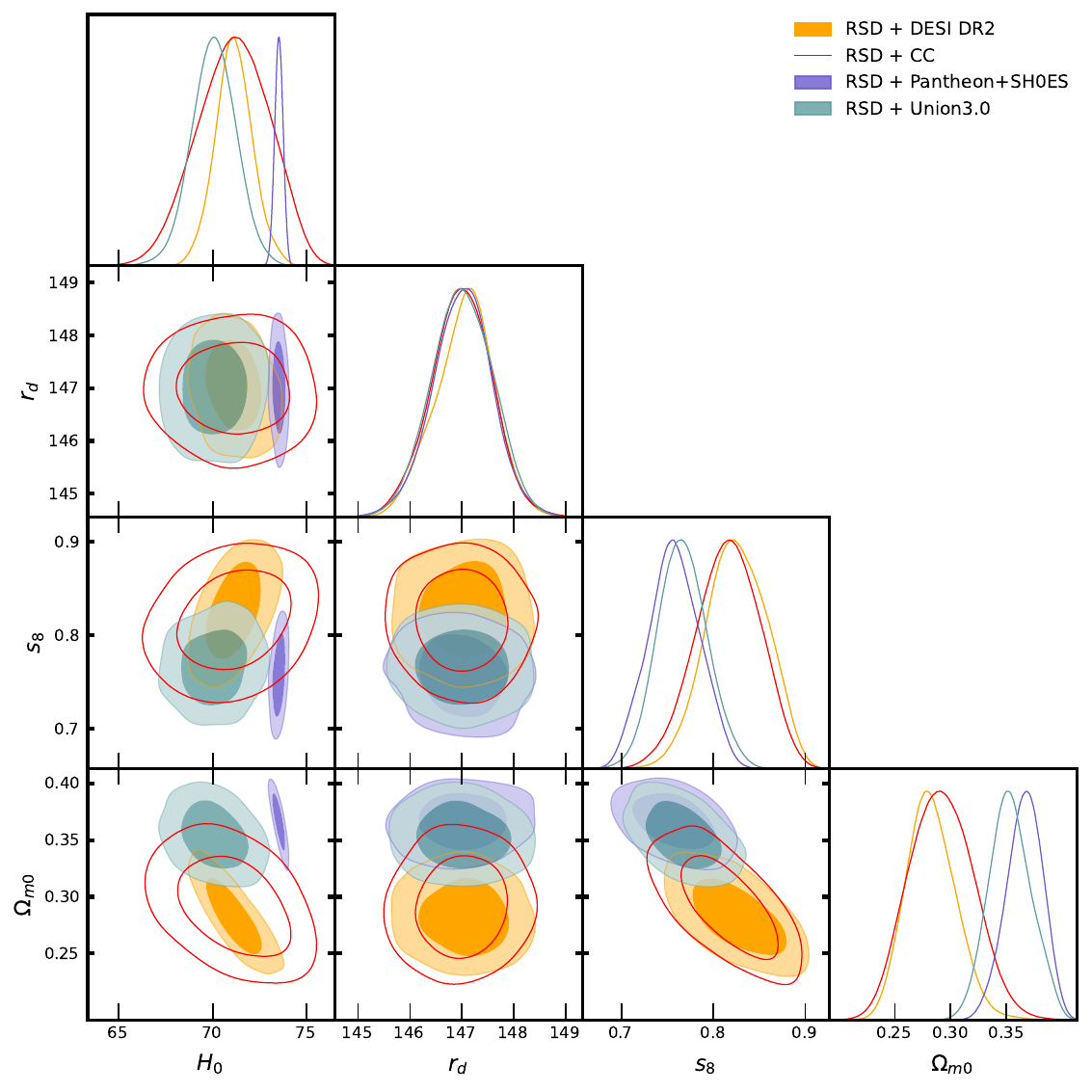}
    \caption{Posterior constraints on the EPN cosmological parameters from different dataset combinations. Contours show the $1\sigma$ and $2\sigma$ confidence regions for the joint posterior distributions.}
    \label{fig:contours}
\end{figure}

\begin{figure}
    \centering
    \includegraphics[width=0.48\linewidth]{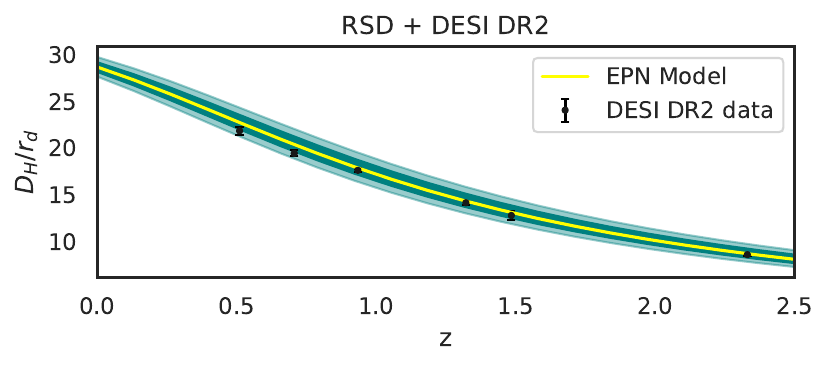}
    \includegraphics[width=0.48\linewidth]{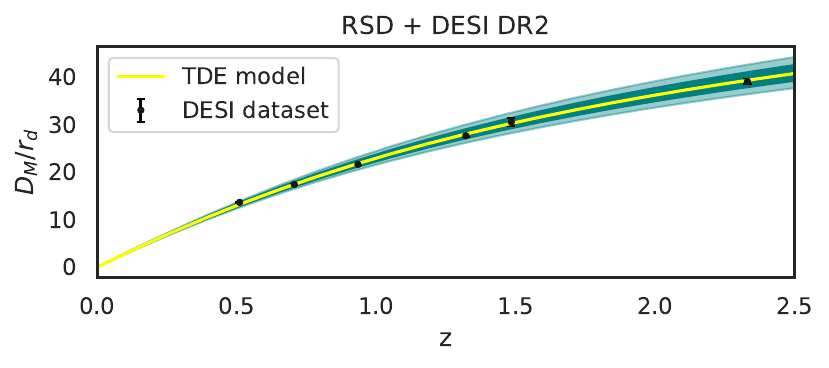}\\
    \includegraphics[width=0.48\linewidth]{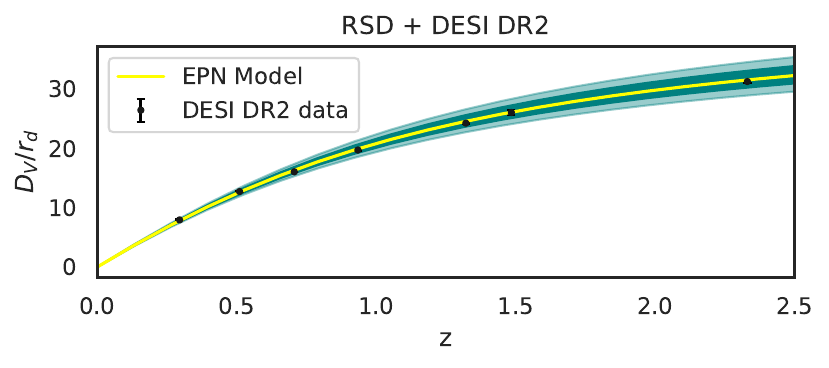}
    \includegraphics[width=0.48\linewidth]{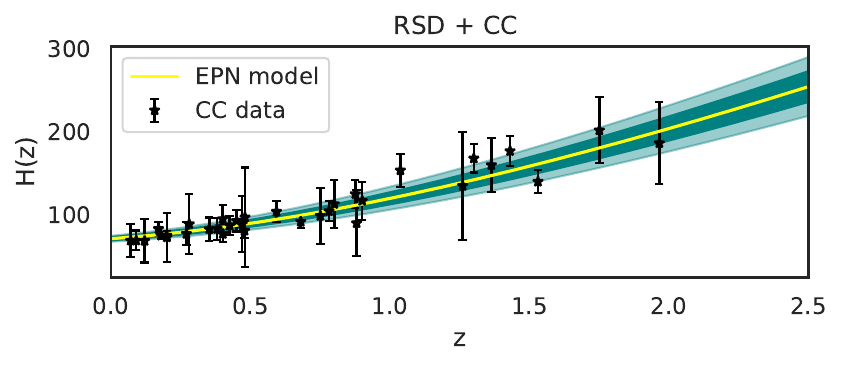}\\
    \includegraphics[width=0.48\linewidth]{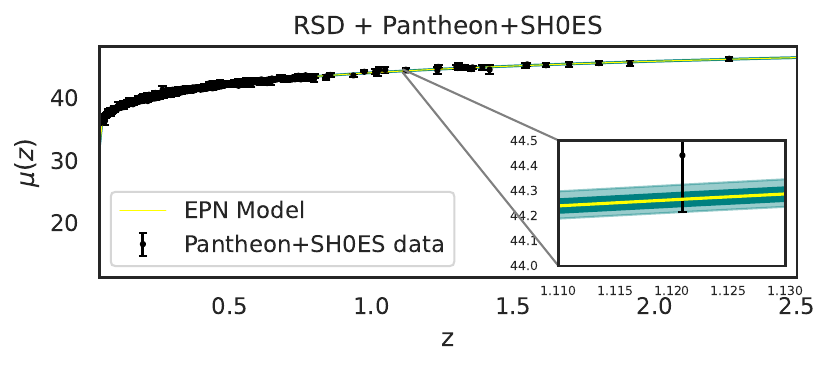}
    \includegraphics[width=0.48\linewidth]{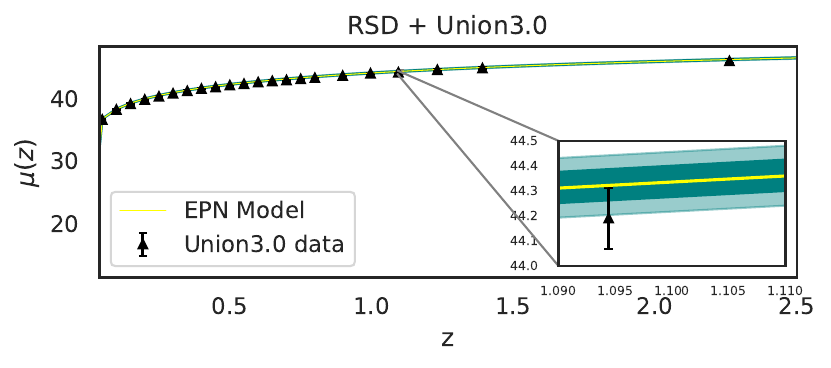}\\
    \includegraphics[width=0.48\linewidth]{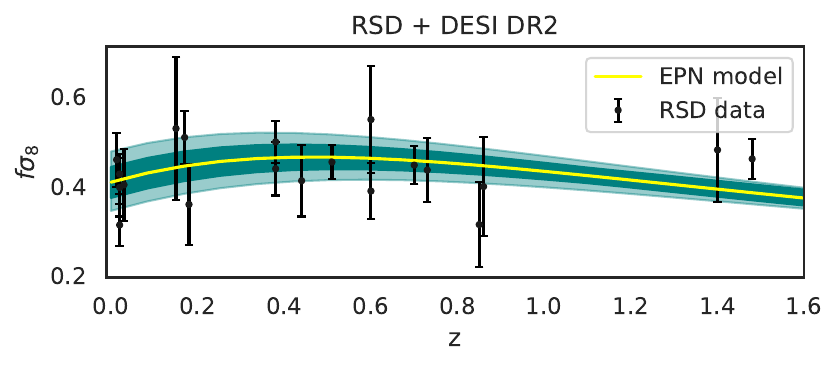}
    \includegraphics[width=0.48\linewidth]{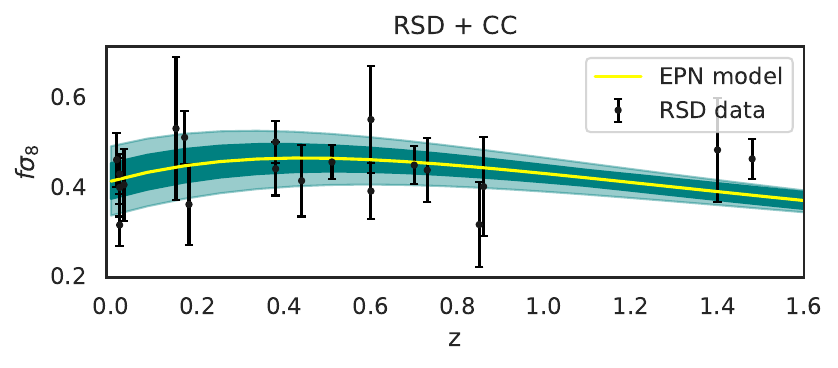}\\
    \includegraphics[width=0.48\linewidth]{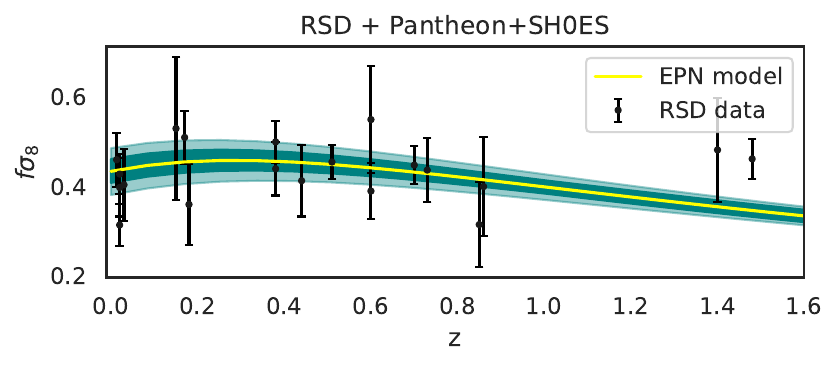}
    \includegraphics[width=0.48\linewidth]{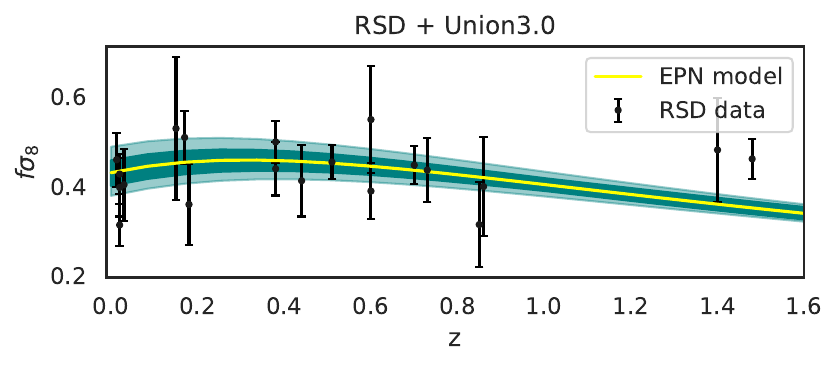}
    \caption{(Top row) BAO measurements from DESI DR2: $D_H/r_d$, $D_M/r_d$, and $D_V/r_d$ (Second row left) as functions of redshift, compared with EPN predictions. (Second row right) Hubble parameter measurements from cosmic chronometers and (third row) supernova distance moduli from Pantheon+SH0ES and Union3.0 (left and right respectively). (Bottom two rows) Growth rate measurements $f\sigma_8(z)$ compared with EPN predictions for different dataset combinations with RSD. In all panels, solid curves show the best-fit EPN predictions, with shaded regions indicating theoretical uncertainties.}
    \label{fig:observations}
\end{figure}

\begin{table}[ht]
\centering
\caption{Parameter constraints from different dataset combinations in the EPN framework. Values shown are the posterior means with their $1\sigma$ uncertainties.}
\label{tab:parameter_constraints}
\begin{tabular}{lcccc}
\toprule
Dataset Combination & $H_0$ [km/s/Mpc] & $r_d$ [Mpc] & $S_8$ & $\Omega_{m0}$ \\
\midrule
RSD + DESI DR2 & $71.1 \pm 1.01$ & $147.06 ^{+0.61}_{-0.51}$ & $0.826^{+0.035}_{-0.032}$ & $0.353^{+0.020}_{-0.024}$ \\
RSD + CC & $71.11 \pm 1.9$ & $146.99 \pm 0.58$ & $0.816 \pm 0.035$ & $0.292 \pm 0.028$ \\
RSD + Pantheon+SH0ES & $73.52 \pm 0.22$ & $147.01 \pm 0.56$ & $0.757 \pm 0.028$ & $0.366^{+0.017}_{-0.015}$ \\
RSD + Union3.0 & $70.1 \pm 1.2$ & $147.02 \pm 0.59$ & $0.767 \pm 0.027$ & $0.353^{+0.018}_{-0.021}$ \\
\bottomrule
\end{tabular}
\end{table}

\section{Results and Analysis}\label{sec:results}

Based on our comprehensive Bayesian analysis of the EPN framework using multiple independent cosmological datasets, we have obtained robust constraints on the key cosmological parameters. The parameter estimates, summarized in Table~\ref{tab:parameter_constraints}, and the corresponding model fits to the observational data reveal a consistent and well-behaved cosmological picture within the EPN paradigm. The Hubble constant $H_0$ varies systematically across dataset combinations, with values ranging from $70.1 \pm 1.2~\unit{km\,s^{-1}\,Mpc^{-1}}$ for RSD combined with Union3.0 to $73.52 \pm 0.22~\unit{km\,s^{-1}\,Mpc^{-1}}$ for RSD with Pantheon+SH0ES. The EPN framework accommodates this range naturally through its modified expansion history, as evidenced by the excellent fits to $H(z)$ data from CC across $0 < z < 2.5$ (see Fig.~\ref{fig:observations}). The predicted expansion evolution smoothly interpolates between early and late universe measurements, with the theoretical curve passing through all observational points within their uncertainties, demonstrating the framework's ability to describe the cosmic expansion across the full observable redshift range. The sound horizon at drag epoch $r_d$ shows remarkable consistency, with all dataset combinations converging around $\approx147$. This uniformity underscores the theoretical robustness of the EPN framework in preserving fundamental cosmological scales. The excellent agreement between EPN predictions and DESI DR2 BAO measurements for $D_V/r_d$, $D_M/r_d$, and $D_H/r_d$ across multiple redshift bins ($0.5 < z < 2.5$) further validates this consistency, with theoretical curves accurately tracing the observational data points throughout the DESI BAO redshift range.

The structure formation parameters show moderate variations that align with the specific characteristics of each dataset. The growth rate evolution plots $f\sigma_8(z)$ in Fig.~\ref{fig:observations} demonstrate that the EPN framework provides excellent fits to RSD data across all combinations, successfully capturing the redshift evolution of cosmic structure formation. The theoretical predictions match the observed transition from slower growth at high redshifts to late-time growth suppression, with all curves lying comfortably within observational uncertainties. The supernova distance modulus plots for both Union3.0 and Pantheon+SH0ES show that the EPN framework provides precise fits to Type Ia supernova data. The predicted distance-redshift relations closely follow the observational points across all redshifts, with slight variations in preferred $H_0$ values reflecting the different calibration approaches of each supernova compilation. Both datasets find comfortable accommodation within the EPN parameter space, demonstrating the framework's flexibility in describing luminosity distance measurements (see Fig.~\ref{fig:observations}). % The matter density parameter $\Omega_{m0}$ shows variations that correspond to the specific redshift coverage and systematic characteristics of each dataset. RSD combined with cosmic chronometers favors $\Omega_{m0} = 0.292 \pm 0.028$, while combinations with BAO and supernovae prefer values around $0.35$. These differences reflect the complementary nature of cosmological probes rather than theoretical inconsistencies, with the EPN framework successfully accommodating the full range through its modified expansion dynamics. Collectively, these results demonstrate that the EPN framework provides excellent fits to all major cosmological probes while maintaining theoretical consistency. 

\section{Discussion and Conclusions}\label{sec:conclusion}

We have presented a comprehensive analysis of cosmological perturbations and structure formation within the EPN framework, establishing for the first time the complete scalar perturbation system for this vector-tensor theory. Our work demonstrates that EPN cosmology, characterized by its algebraic constraint~\eqref{eq:constraint_epn} and powered-Hubble expansion history~\eqref{eq:HubbleAlgebraic}, provides a theoretically consistent and observationally viable alternative to the standard $\Lambda$CDM paradigm.

The key theoretical insight emerging from our analysis is that the EPN framework modifies cosmic evolution through two distinct but interrelated mechanisms. First, at the background level, the algebraic constraint between the vector field and Hubble parameter generates an effective dark-energy sector whose density scales as $ {\rho}_{\rm EPN}\propto H^{-2/3}$. This non-canonical scaling introduces characteristic deviations from $\Lambda$CDM at intermediate redshifts while preserving standard behavior in both early and late asymptotic limits~\citep{Anagnostopoulos:2023pvi}. Second, at the perturbative level, the EPN sector contributes as an effective fluid with anisotropic stress ($\Phi\neq\Psi$), coupling the surviving scalar vector mode to the metric potentials through Eqs.~\eqref{eq:E1}-\eqref{eq:E4} while maintaining ghost-free stability through carefully designed interaction structures. The persistence of the background constraint at linear order ensures that only one additional scalar degree of freedom propagates, preserving the theory's theoretical consistency.

From an observational perspective, our multi-probe Bayesian analysis reveals that the EPN framework successfully accommodates measurements from DESI BAO, Pantheon+ and Union3 supernovae, CC, and RSD. The parameter constraints in Table~\ref{tab:parameter_constraints} show systematic variations across dataset combinations that reflect the complementary nature of cosmological probes rather than theoretical inconsistencies. The excellent fits to $f\sigma_8(z)$ growth data, shown in Fig.~\ref{fig:observations}, confirm that the minimally coupled EPN scenario maintains viable structure formation while modifying the background expansion.

The underlying physics of the EPN framework offers several distinctive features that merit further exploration. Unlike scalar-tensor theories where additional fields typically introduce new dynamical degrees of freedom, the EPN vector field is constrained by algebraic relations that eliminate unwanted modes while preserving nontrivial self-interactions. This constrained structure provides theoretical control over stability conditions (Appendix~\ref{app:coeffs-stability}) while allowing for rich phenomenological behavior through different parametrization families. The emergence of anisotropic stress in the perturbation sector represents a potentially observable signature distinguishing EPN from $\Lambda$CDM, with implications for weak gravitational lensing and integrated Sachs-Wolfe measurements~\citep{Saltas:2014dha,Tutusaus:2022cab,Sobral-Blanco:2021cks}.

Several promising directions for future research emerge from our analysis. First, extending the perturbation analysis to include tensor modes and vector perturbations would provide a complete picture of gravitational wave propagation and cosmic microwave background anisotropies in the EPN framework. Second, investigating non-minimal matter-vector couplings could reveal richer phenomenology in the structure growth sector, potentially altering the effective gravitational coupling $G_{\rm eff}$ on cosmological scales. Third, application to high-redshift probes such as Lyman-$\alpha$ forest measurements and CMB lensing would test the framework's predictions across cosmic history. Finally, a detailed comparison with other modified gravity scenarios, including Horndeski theories and massive gravity, would help situate EPN within the broader landscape of beyond-GR frameworks.

\section*{Data availability} 
There are no new data associated with this article.

\section*{Acknowledgments}

The research was supported by the Ministry of Higher Education (MoHE), through the Fundamental Research Grant Scheme (FRGS/1/2023/STG07/UM/02/3, project no.: FP074-2023). NSK acknowledges the Inter-University Centre for Astronomy and Astrophysics (IUCAA) for the support and opportunity extended through its Visitor Programme.

\appendix

\section{Perturbation Coefficients and Stability Conditions}
\label{app:coeffs-stability}

For completeness, we collect here the explicit expressions entering the scalar perturbation sector of the EPN theory. These coefficients appear in the dynamical vector--mode equation~\eqref{eq:V2} and in the effective-fluid perturbations $(\delta \rho_{\rm EPN},\delta  p_{\rm EPN},  q_{\rm EPN}, \pi_{\rm EPN})$ used in Eqs.~\eqref{eq:E1}--\eqref{eq:E4} in the main text. All quantities are evaluated on the homogeneous background solution $(H,\phi,X)$.

\subsection{Kinetic operator and sound speed of the propagating scalar mode}

The quadratic action for the surviving EPN scalar mode can be written in the schematic form
\begin{equation}
S^{(2)}_{\rm EPN}
=\frac{1}{2}\int d^3k\,dt\,a^3
\left[
\mathcal{K}_{\rm EPN}(a)\,\dot{\hat{\chi}}^{\,2}
-\mathcal{G}_{\rm EPN}(a,k)\,\frac{k^2}{a^2}\hat{\chi}^{\,2}
-\mathcal{M}_{\rm EPN}(a)\,\hat{\chi}^{\,2}
\right],
\label{eq:app-quad-action}
\end{equation}
where $\mathcal{K}_{\rm EPN}>0$ ensures the absence of ghosts, while the effective sound speed is defined as
\begin{equation}
c_s^2(a)
\equiv
\frac{\mathcal{G}_{\rm EPN}(a,k)}{\mathcal{K}_{\rm EPN}(a)}
\quad (\text{sub-horizon limit } k\gg aH).
\label{eq:cs2def}
\end{equation}

In terms of the interaction functions $\alpha_0(X)$, $\alpha_1(X)$, and $d_1(X)$ and their derivatives, the kinetic operator reads
\begin{equation}
\mathcal{K}_{\rm EPN}(a)
=
\frac{\phi^2}{\Lambda^2}
\left[
\alpha_{0,XX}
+3(\alpha_{1,XX}+d_{1,XX})\frac{H\phi}{\Lambda^2}
+2(\alpha_{1,X}+d_{1,X})\frac{H}{\Lambda^2}
\right],
\label{eq:Kcoeff-app}
\end{equation}
while the spatial-gradient coefficient takes the form
\begin{equation}
\mathcal{G}_{\rm EPN}(a)
=
(\alpha_{1,X}+d_{1,X})
\left[
1+\frac{\dot{H}}{H^2}
+\frac{\dot{\phi}}{H\phi}
\right]
+\frac{\phi^2}{3\Lambda^2}
\left(
\alpha_{1,XX}+d_{1,XX}
\right).
\label{eq:Gcoeff-app}
\end{equation}
The effective mass contribution is given by
\begin{equation}
\mathcal{M}_{\rm EPN}(a)
=
H^2\Big[
\alpha_{0,X}
+(\alpha_{1,X}+d_{1,X})
\left(
3+\frac{\dot{\phi}}{H\phi}
\right)
\Big].
\label{eq:Mcoeff-app}
\end{equation}

The stability-screening conditions employed in the main text correspond to
\begin{equation}
\mathcal{K}_{\rm EPN}(a)>0,
\qquad
c_s^2(a)=\frac{\mathcal{G}_{\rm EPN}(a)}{\mathcal{K}_{\rm EPN}(a)}>0 ,
\label{eq:stability-conditions}
\end{equation}
and are evaluated numerically along the background solution.

\subsection{Coefficients in the dynamical perturbation equation}

The coefficients appearing in Eq.~\eqref{eq:V2} may be written as
\begin{align}
\mathcal{D}(a)
&=a^3\mathcal{K}_{\rm EPN}(a), \label{eq:Dcoeff} \\
\mathcal{E}(a)
&=\frac{d}{dt}\!\left[a^3\mathcal{K}_{\rm EPN}(a)\right],
\label{eq:Ecoeff}\\
\mathcal{F}(a)
&=a^3\mathcal{M}_{\rm EPN}(a),
\label{eq:Fcoeff}
\end{align}
while the source couplings to the metric potentials are given by
\begin{equation}
\mathcal{S}_\Phi(a)
=
a^3\phi H(\alpha_{1,X}+d_{1,X}),
\qquad
\mathcal{S}_\Psi(a)
=
-2a^3H(\alpha_{1,X}+d_{1,X}),
\label{eq:sourcepotcoeff}
\end{equation}
and the scale-dependent contribution reads
\begin{equation}
\mathcal{S}_k(a)
=
a^3(\alpha_{1,X}+d_{1,X}).
\label{eq:Skcoeff}
\end{equation}

These expressions are used in all numerical calculations reported in the main text.

%\section{Parametrization Families for the EPN Interaction Functions}
%\label{app:param-families}

%This appendix summarizes the functional forms of the parametrization families introduced in the main text, together with the associated parameter ranges and theoretical constraints.

%\subsection{Family~I: Power--law parametrization (baseline class)}

%The baseline class assumes power--law dependence of the interaction %functions on the invariant $X$,
%\begin{equation}
%\alpha_0(X)=A_0\left(\frac{X}{X_0}\right)^{n_0},\qquad
%\alpha_1(X)=A_1\left(\frac{X}{X_0}\right)^{n_1},\qquad
%d_1(X)=D_1\left(\frac{X}{X_0}\right)^{m_1},
%\label{eq:powerlawfamily}
%\end{equation}
%where $X_0$ denotes the present-day value of $X$. The free parameters
%\[
%\{A_0,A_1,D_1,n_0,n_1,m_1\}
%\]
%are constrained by the background algebraic relation~\eqref{eq:constraint_epn} and by the stability conditions~\eqref{eq:stability-conditions}. This family provides a minimal theory-motivated benchmark with simple scaling behaviour.

\bibliography{main}
\end{document}